\documentclass[11pt]{article}
\usepackage[dvips]{graphicx}  
\begin{document}

\title{\LARGE \bf MICHELSON-MORLEY EXPERIMENTS REVISITED and the COSMIC BACKGROUND
RADIATION PREFERRED FRAME }  
\author{{Reginald T. Cahill and Kirsty Kitto}\\
  {School of Chemistry, Physics and Earth Sciences}\\{ Flinders University
}\\ { GPO Box
2100, Adelaide 5001, Australia }\\{(Reg.Cahill@flinders.edu.au)}
\\{(Submitted to {\it Nature})}}

\date{}
\maketitle

\begin{abstract}
{\it We report a simple re-analysis of the old results (1887) from the Michelson-Morley interferometer
experiment that was designed to detect absolute motion.  We  build upon
 a recent (1998) re-analysis of the original data  by M\'{u}nera, which 
revealed a small but significant effect after allowing for several systematic errors in the original analysis.
 The further re-analysis here reveals that a genuine
effect of absolute motion is expected, in what is essentially a quantum interference experiment,  but only if the
photons travel in the interferometer at speeds
$V<c$.  This is the case if the interferometer  operates in a dielectric, such as air, as was the
case, incidently, of the Michelson-Morley experiment.  The re-analysis here of the Michelson-Morley experimental
data, correcting for the refractive index effect of the air,  reveals an absolute
speed of the Earth of $v=359\pm54$ km/s, which is in excellent agreement with the speed of
$v=365\pm18$ km/s determined from the dipole fit, in 1991, to the NASA  COBE satellite Cosmic Background
Radiation (CBR) observations.  Other experiments where the interferometers operated in air (Miller 1925,1933) or helium
(Illingworth 1927)  give similar results  when re-analysed.  These experimental results refute Einstein's assertion that
absolute motion through space has no meaning. This re-analysis was motivated by  developments in a new information-theoretic
modelling of reality, known as Process Physics\footnote{{\bf Process Physics} Web Page:
http://www.socpes.flinders.edu.au/people/rcahill/processphysics.html}, in which the Einstein Special and General Relativity
formalisms arise as consequences of an emergent quantum-foam explanation for space. This amounts to  an absolute and 
preferred frame of reference, in what is a unified quantum theory of space, gravity and  matter.  So Einstein's theoretical
phenomenology survives, and is now explained, even though  he used a false premise in his derivation, but his spacetime
construct is now seen to have no ontological significance. We predict that  new interferometer experiments, operating in
dielectric mode,  will reveal an inward velocity component of magnitude $11.2$km/s due to the quantum-foam flow caused by the
matter of the Earth.  This is a fundamental test of the quantum theory of gravity that has emerged from Process Physics.}

\end{abstract}

\vspace{3mm}

COBE, preferred frame, quantum
foam, quantum gravity, process physics, aether. 

\vspace{3mm}

\newpage

One key unchallenged folklore in physics is that the Michelson interferometer laboratory experiment of
1881 \cite{M}, and repeated by Michelson and Morley in 1987 \cite{MM}, that were designed to detect absolute
motion, gave a null result,  vindicating Einstein's assumption that absolute motion (motion relative to space
itself) has no meaning; it is {\it in principle} not detectable in a laboratory situation. Motion of objects
is always relative to other objects, according to Einstein. Using this assumption Einstein went on to
construct the Special and General Theory of Relativity, which uses the notion of spacetime to avoid any notion
of absolute space.  Of course Einstein's formalism has been abundantly confirmed both by the extensive use of
the special theory in particle physics experiments and theory, and by the general theory in various
experimental and observational situations.  Nevertheless we report here experimental evidence that absolute
motion is detectable in  laboratory experiments, such as those done by Michelson and Morley and others, but that this
requires a re-analysis of the operation of their interferometer, as reported herein. This analysis
leads to a speed which agrees with that found from the  NASA COBE satellite observations on analysing the dipole
anisotropy of the Cosmic Background Radiation. These two speed observations,   104 years apart,  agree within
experimental errors. Together these results show that absolute motion has been detected.   New
interferometer experiments are needed to confirm that  the direction  of that motion is the same as the direction
discovered by the COBE mission. These results are profoundly significant to our understanding of reality. It
follows from recent work that these experimental outcomes will not be in conflict with the Einstein phenomenology,
but require a major re-assessment of what that phenomenology describes  \cite{RC02}. 

Recent developments in the new
information-theoretic modelling of reality, known as {\it Process Physics} \cite{RC02, RC01}, indicate that the
Einstein Special and General Relativity formalisms arise as a consequence of an emergent quantum-foam
explanation for space, but with this quantum foam amounting to  an absolute and  preferred frame of reference,
in what is a unified quantum theory of space, gravity and matter.  In \cite{RC02} it was shown how the
general theory account of gravity arises from an amalgam of two distinct quantum foam effects, one being the
effective diffusion of the quantum foam towards `matter' that acts as a `sink', together with the classical
measurement protocol based on the apparent invariance of the speed of light, which uses  the radar method
to retrospectively assign spacetime coordinates to distant events, that is, spacetime is in fact a
historical record of reality, and not reality itself.   Here the invariance of the speed of light is caused by
the genuine dynamical effect  of the quantum foam on moving rods and clocks, as suggested long ago by
Fitzgerald and Lorentz to explain the, here now disputed,  null result of the Michelson-Morley experiment. Observers
adopting this phenomenological property of light within the classical measurement protocol will find, according
to Process Physics, that  the formalism of the General Theory of Relativity describes their historical records
which have the form of a spacetime geometrical construct, but that this spacetime construct  has no ontological
significance. In this construct the modelling of time by geometry is entirely appropriate; it is essentially a
pagination of the historical record.  The resolution of the confusion between the geometrical modelling of time
in historical records and time as  an actual process has now been achieved within the new {\it Process
Physics}. The older but still current {\it Non-Process Physics} modelling of reality, which has been with us
since  Galileo  and Newton introduced the geometrical modelling of time,  is now superseded. 

An implication of these developments in {\it Process Physics} is that it should be possible {\it in principle}
to overcome the classical measurement protocol effects by using the long-known nonlocality of quantum processes
which should be sensitive to absolute motion  through this quantum foam.  Indeed ever since the Aspect
\cite{Aspect} experiment showed a violation of the Bell inequalities, in the context of the Einstein-Podolsky-Rosen (EPR)
 nonlocal effects it has been understood that quantum collapse events caused by quantum detectors appear to
require a preferred frame, and so appear to be in conflict with Einstein's assumption.  The work of Hardy
\cite{Hardy} and Percival \cite{Percival}  suggested that `double-EPR' experiments  could reveal an
absolute frame, and hence absolute motion, though such experiments for this purpose would be extremely
difficult.  However the old Michelson-Morley interferometer experiment is actually a quantum interferometer,
that is,  in principle it could be done with one photon at a time, and we inquired why it is believed that
it was unable to detect a preferred frame, and so absolute motion. In fact it can do so.  

As described in
Fig.1 the  beamsplitter/mirror
$A$ sends a photon $\psi(t)$ into a superposition
$\psi(t)=\psi_1(t)+\psi_2(t)$, with each component travelling in different arms of the interferometer, until
they are recombined in the quantum detector which results in a localisation process, and one spot in the
detector is produced.  Repeating with many photons, reveals that the interference between $\psi_1$ and
$\psi_2$ at the detector results in fringes.  Before  the quantum theory (and in particular before the new
quantum foam physics)  Michelson   designed the interferometer with the idea that any motion
through absolute space (then called the luminiferous aether) would result in different path lengths for light
waves in the two arms, which  would show as a shift in the fringe. 
Of course as an experimental expediency one rotates the apparatus through $90^o$ so that the roles of the two
arms are interchanged, this rotation then should result in a shift of the fringes, which is an effect more easily observed. 
However Michelson and Morley reported  a null result, despite a small effect actually being seen.  Fitzgerald and then Lorentz
then offered an explanation for the null result, namely that the failure to get an effect was caused by the actual
contraction of the arm moving lengthwise through the absolute space, as we show below. However long ago it
was decided by physicists that the more elaborate Einstein  explanation in terms of spacetime transformations
was superior to this dynamical explanation.   

In reviewing the operation of the Michelson-Morley interferometer (see below) it was noticed that the
Fitzgerald-Lorentz contraction explanation only implies a null effect if the experiment is performed in
vacuum. In air, in which photons travel slightly slower than in vacuum, there should be a small fringe shift
effect when the apparatus is rotated, even after taking account of the Fitzgerald-Lorentz contraction.   This
appears to have gone unnoticed.  As well M\'{u}nera
\cite{Munera} in 1998 has corrected the original results for various systematic errors, and shown that these
interferometer experiments do indeed reveal small but significant non-null results, but  as expected now, only
when they are operated  in a dielectric.

We have applied the elementary correction required for the effects of the air, or in one case helium,  to
various interferometer  results as corrected by M\'{u}nera for other  systematic errors.   The correction
is in fact large, being some two orders of magnitude. Applied to one set of data the Michelson-Morley experiment
now results in a  speed of  $v=359\pm54$ km/s.  Even more amazing is the excellent
agreement of this speed with the speed of $v=365\pm18$ km/s determined from the dipole fit, in 1991, to the
NASA  COBE satellite Cosmic Background Radiation observations \cite{Smoot}.  

We now indicate the proper understanding of the operation of the Michelson-Morley interferometers when operated
in a dielectric, and then report the corrected results, as shown in Fig.2. 
The two arms are constructed to have the same lengths  when they are physically parallel to each other.
For convenience assume that the value $L$ of this length   refers to the lengths when at rest in the quantum
foam. The Fitzgerald-Lorentz effect is that the arm $AB$  parallel to the direction of motion is shortened to
\begin{equation}
L_{\parallel}=L\sqrt{1-\frac{v^2}{c^2}}
\end{equation}
by motion through the quantum foam. 
\vspace{15mm}
\begin{figure}[h]
\hspace{30mm}
\setlength{\unitlength}{1.0mm}
\begin{picture}(0,20)
\thicklines
\put(-10,0){\line(1,0){50}}
\put(-5,0){\vector(1,0){5}}
\put(40,-1){\line(-1,0){29.2}}
\put(15,0){\vector(1,0){5}}
\put(30,-1){\vector(-1,0){5}}
\put(10,0){\line(0,1){30}}
\put(10,5){\vector(0,1){5}}
\put(11,25){\vector(0,-1){5}}
\put(11,30){\line(0,-1){38}}
\put(11,-2){\vector(0,-1){5}}
\put(8.0,-2){\line(1,1){5}}
\put(9.0,-2.9){\line(1,1){5}}
\put(6.5,30){\line(1,0){8}}
\put(40,-4.5){\line(0,1){8}}
\put(5,12){ $L$}
\put(4,-5){ $A$}
\put(35,-5){ $B$}
\put(25,-5){ $L$}
\put(12,26){ $C$}
\put(9,-8){\line(1,0){5}}
\put(9,-9){\line(1,0){5}}
\put(14,-9){\line(0,1){1}}
\put(9,-9){\line(0,1){1}}
\put(15,-9){ $D$}

\put(50,0){\line(1,0){50}}
\put(55,0){\vector(1,0){5}}
\put(73,0){\vector(1,0){5}}
\put(85,0){\vector(1,0){5}}
\put(90,15){\vector(1,0){5}}
\put(100,-1){\vector(-1,0){5}}
\put(100,-4.5){\line(0,1){8}}
\put(68.5,-1.5){\line(1,1){4}}
\put(69.3,-2.0){\line(1,1){4}}
\put(70,0){\line(1,4){7.5}}
\put(70,0){\vector(1,4){3.5}}
\put(77.5,30){\line(1,-4){7.7}}
\put(77.5,30){\vector(1,-4){5}}
\put(73.5,30){\line(1,0){8}}
\put(83.3,-1.5){\line(1,1){4}}
\put(84.0,-2.0){\line(1,1){4}}
\put(100,-1){\line(-1,0){14.9}}
\put(67,-5){ $A_1$}
\put(82,-5){ $A_2$}
\put(95,-5){ $B$}
\put(79,26){ $C$}
\put(90,16){ $v$}
\put(-8,8){(a)}
\put(55,8){(b)}

\end{picture}
\vspace{10mm}
\caption{\small{Schematic diagrams of the
Micheslon Interferometer, with
beamsplitter/mirror at $A$ and mirrors at $B$ and
$C$, on equal length arms when parallel,
from $A$. $D$ is a quantum detector (not drawn in (b)) that causes localisation 
of the photon state by a collapse process. In (a)
the interferometer is at rest in the quantum foam. In (b) the
interferometer is moving with speed $v$
relative to the quantum foam in the direction
indicated. Interference
fringes are observed at the quantum detector $D$. 
  If the interferometer is
rotated in the plane  through $90^o$, the
roles of arms
$AC$ and $AB$ are interchanged, and during
the rotation shifts of the fringes are seen
in the case of absolute motion, but only if the apparatus operates in a dielectric.  By counting
fringe changes the speed $v$ may be
determined.}}
\end{figure}
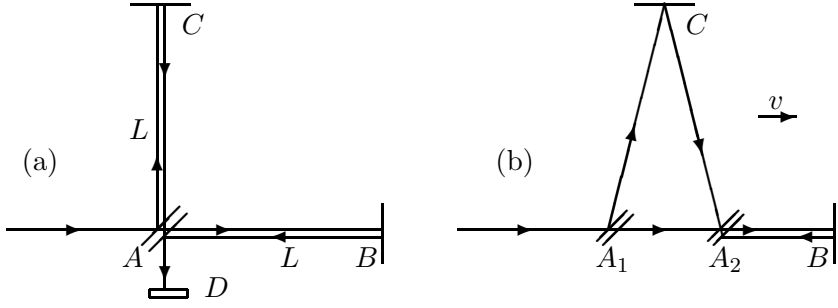

Following Fig.(1), we consider the case when  the apparatus is moving at speed $v$ through the quantum foam,
and  that the photon states  travel at speed $V=c/n$ relative to the quantum foam, where $n$ is the refractive index of the gas
and
$c$ is the speed of light, in vacuum,  relative to the quantum foam.  Let the time taken for $\psi_1$ to travel from
$A\rightarrow B$ be
$t_{AB}$ and that from $B\rightarrow A$ be $t_{BA}$. 
In moving from the beamsplitter at $A$ to $B$, the photon state $\psi_1$ must travel an extra  distance 
because
 the mirror $B$ travels a distance $vt_{AB}$ in this time, thus the total distance that must be traversed  is
\begin{equation}
Vt_{AB}=L_{\parallel}+vt_{AB}.
\end{equation}
Similarly, on returning from $B$ to $A$, the photon state $\psi_1$ must travel the distance
\begin{equation}
Vt_{BA}=L_{\parallel}-vt_{BA}.
\end{equation}
Hence the total time $t_{ABA}$ taken for $\psi_1$  to travel from $A\rightarrow B \rightarrow A$ is given
by
\begin{eqnarray}\label{eq:ABA}
t_{ABA}=t_{AB}+t_{BA}&=&\frac{L_{\parallel}}{V-v}+\frac{L_{\parallel}}{V+v}\\
&=&\frac{L_{\parallel}(V+v)+L_{\parallel}(V-v)}{V^2-v^2}\\
&=&\frac{2LV\sqrt{1-\displaystyle\frac{v^2}{c^2}}}{V^2-v^2}.
\end{eqnarray}
Now, assuming that the time taken for the photon state $\psi_2$ to travel from $A\rightarrow C$ is $t_{AC}$,
but that the apparatus travels a distance $vt_{AC}$ in that time, we can use the Pythagoras theorem:
\begin{equation}
\left(Vt_{AC}\right)^2=L^2+\left(vt_{AC}\right)^2
\end{equation}
which gives
\begin{equation}
t_{AC}=\frac{L}{\sqrt{V^2-v^2}},
\end{equation}
and including the return trip ($C\rightarrow A,  t_{CA}=t_{AC}, t_{ACA}=t_{AC}+t_{CA}$) results in 
\begin{equation}\label{eq:ACA}
t_{ACA}=\frac{2L}{\sqrt{V^2-v^2}},
\end{equation}
giving finally for the time difference for the two arms
\begin{equation}
 \Delta t= \frac{2LV\sqrt{1-\displaystyle\frac{v^2}{c^2}}}{V^2-v^2}-\frac{2L}{\sqrt{V^2-v^2}}.
\end{equation}
Now trivially $\Delta t =0$  if $v=0$, but  also $\Delta t =0$ when $v\neq 0$ but only if $V=c$.  This then 
  would  result in a null result on rotating the apparatus.  Hence the null result of the Michelson-Morley
apparatus is only for the special case of photons travelling in vacuum for which $V=c$, as confirmed by
the modern vacuum interferometer experiment of Brillet and Hall
\cite{BH}, which in-effect confirms (1).   However if the apparatus is immersed in a gas then $V<c$ and a non-null
effect is expected on rotating the apparatus, since now  $\Delta t \neq 0$.  It is essential then in analysing data
to correct for this refractive index effect.  Putting
$V=c/n$ in (10) we find, for $v << V$ and  when $n \approx 1$, that 
\begin{equation}
\Delta t= -L(n-1)\frac{v^2}{c^3}.
\end{equation}
 However if the data is analysed not using the
Fitzgerald-Lorentz contraction (1), then, as done in the old analyses,   the estimated time difference is 
\begin{equation}
\Delta t = \frac{2LV}{V^2-v^2}-\frac{2L}{\sqrt{V^2-v^2}},
\end{equation}
which again for $v << V$ and $n \approx 1$, gives 
\begin{equation}
\Delta t =- L\frac{v^2}{c^3}.
\end{equation}
The value of $\Delta t$ is deduced from analysing the fringe shifts, and then    the speed $v_{MM}$ (in previous
Michelson-Morley analyses) has been extracted  using (13), instead of the correct form (11).  $\Delta t$
 is typically of order $10^{-15}s$.   However it is very easy to correct for this oversight.  From (11)
and (13) we obtain, for the corrected absolute speed through the quantum-foam $v_{QF}$, 
\begin{equation}
v_{QF}=\frac{v_{MM}}{\sqrt{n-1}}.
\end{equation}
Note that for air at STP
$n=1.00029$, while for helium at STP $n=1.000036$, and so the correction factor of $1/\sqrt{n-1}$ is large.
The corrected speeds $v_{QF}$ for four MM experiments are shown in Fig.2, and compared with the CBR speed
determined from the COBE data  \cite{Smoot}.

M\'{u}nera \cite{Munera} has reviewed these interferometer experiments and uncovered systematic errors  and as well applied
standard statistical tests to the values originally reported.   He has noted that the Michelson-Morley
experiments and subsequent repetitions never were null, and that correcting for invalid inter-session averaging
leads to  even larger non null results.  M\'{u}nera's new results are as follows.  The original Michelson-Morley
data \cite{MM} now gives $v_{MM}=6.22$km/s with a standard deviation on the mean of $0.93$km/s for one set of
noon sessions, while giving $v_{MM}=6.80$km/s with  a standard deviation on the mean of $2.49$km/s for $18^h$
observations.  For Miller \cite{Miller1, Miller2}, from results at Mt Wilson after moving the Morley-Miller
\cite{MorMill} experiment,  the new result is $v_{MM}=8.22$km/s with an upperbound at 95\% CL of $v_{MM}=9.61$km/s and
a lower bound at 95\% CL of $v_{MM}=6.83$km/s. The Illingworth \cite{Illingworth} data (from a helium-filled
inteferometer)  gives smaller values  of $v_{MM}=3.13$km/s with an upperbound at 95\% CL of $v_{MM}=4.17$km/s and a
lower bound at 95\% CL of
$v_{MM}=2.09$km/s.

 Hence the Michelson-Morley and Miller values for $v_{MM}$ appear to differ significantly from the
value from Illingworth. But this is now an expected outcome as Illingworth used helium 
(to control temperature variations) instead of air.
 Because of the different refractive indices of air and helium the correction factors are
substantially different ($1/\sqrt{n-1}=58.7$  for air at STP and = 166.7 for helium at STP),  and as shown in
Fig.2 the air and helium filled interferometers  now give comparable results, when corrected.  The use of
helium by Illingworth has turned out to be a fortuitous situation that allows us to confirm the correction-factor
dependence on $n$ in (14).  We have assumed that Illingworth used helium at STP, but if in fact he used helium at 2
atmospheres then his corrected data, in Fig.2, would coincide with the COBE data.  We have also used the
$n$ value for air at STP in correcting the Miller results even though the  experiments were performed at  altitude on Mt.
Wilson. 

\vspace{5mm}
\hspace{15mm}\includegraphics[scale=1.5]{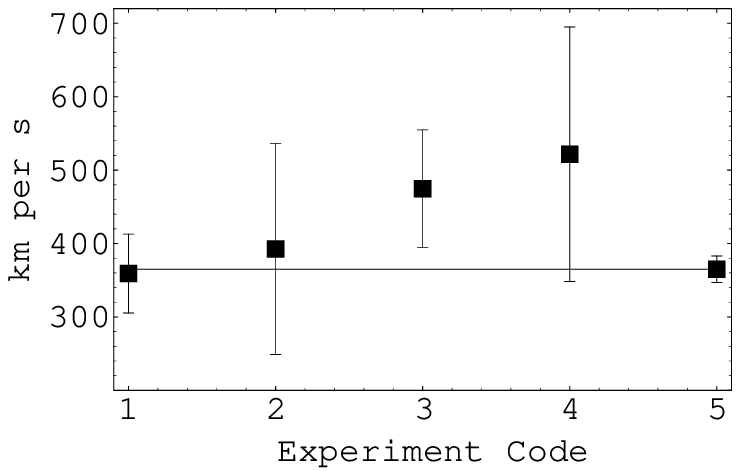}
\begin{figure}[ht]
\caption{\small 
Speeds $v_{QF}$ in km/s determined from various Michelson-Morley experiments: ({\bf 1})
Michelson-Morley (noon observations) \cite{MM}, ({\bf 2}) Michelson-Morley experiments ($18^h$ observations)
\cite{MM}, ({\bf 3}) Miller, Mt. Wilson \cite{Miller1, Miller2}, ({\bf 4}) Illingworth \cite{Illingworth}, and finally
in ({\bf 5}) the speed from the COBE satellite observation of the CBR microwave spectrum dipole term
\cite{Smoot}. The results ({\bf 1})-({\bf 4}) are not corrected for the $\pm30$km/s of the orbital motion of
the earth about the sum, though this correction was made, as well as that for the  satellite orbital speed,
in the case of ({\bf 5}). The agreement between the Michelson-Morley result in ({\bf 1}) and the COBE result
in ({\bf 5}) is striking. The horizontal line at $v=365$km/s is to aid comparisons with the COBE data. } 
\end{figure}

Of course the vacuum experiment by Brillet and Hall \cite{BH} gave a genuine
null outcome as now expected.   The presence of gases in these early experiments, rather than high vacuum, was
of course an experimental expediency, but only because of this can we now realise the full implications of these
long forgotten experiments.  

Hence the old interferometer data  is not only non-null but the extracted speeds agree, within errors,  with the
speed determined from analysis of the CBR dipole component  observed by the COBE mission.  These results demonstrate that
absolute motion has a meaning and is measureable.

A new class of experiment can now be carried out using say a rotateable
fibreoptic laser interferometer, and such quantum-gravity experiments  are
capable of  measuring the absolute speed and direction  of motion of the Earth  through the quantum foam that is space. 
As well the very large gravity-wave laser interferometers could be used in a gas-filled mode rather than a
vacuum mode and, if sufficient temperature control can be achieved, they  could be used as absolute motion
detectors.  
Daily and seasonal changes in
$v_{QF}$ will be seen as 
$\vec{v}_{QF}$ is the vector sum of various contributions due to the  effects of the  dissipative and non-uniform flows
of the quantum foam (which is gravity, \cite{RC02}) by the earth, moon and sun, as well as the overall effect of the motion of
the Earth and the solar system through the CBR determined quantum-foam frame of reference which is, it now turns out, was one of
the main discoveries of the COBE mission.  

  A  task for new interferometers  would be to establish the in-ward flow rate of the quantum foam, predicted to be
$11.2$km/s at the Earth's surface.  This quantum-foam flow is a key test of the quantum theory of gravity that has emerged from
{\it Process Physics} \cite{RC02, RC01,
CKK00, CK99,CK98, MC}, and in which  Newton's gravitational constant $G$ is a measure of the rate at which `quantum matter'
dissipates quantum foam. 

The significance of this new interpretation  of the outcomes of the various  Michelson-Morley experiments can
hardly be overstated, it will change forever how physicists comprehend  reality. 
These results undermine Einstein's assertion that absolute motion has no meaning.  This {\it Process Physics}  supersedes 
the prevailing {\it Non-Process Physics} modelling of reality, which is characterised by the geometrical modelling of time.
 It is remarkable that these  experiments were carried out with such diligence and care so long ago that
their data, when now properly analysed, yield speeds  consistent with those found from  satellite technology 104
years later, and even more so when we understand that some of the experimenters themselves believed they had failed
to detect non-null effects.    
In effect the Michelson interferometer, operating with  dielectric medium, is a
quantum interferometer that performed  the first quantum gravity experiment as it was capable of detecting absolute motion
through the quantum foam that is space.

Again we emphasis  that the Einstein Special and General Relativity is essentially  a model of the potential history of
the universe rather than of the universe itself.  Clearly if the universe is lawful, then so is its history, though the
`laws' for the latter will clearly be different  and, as is well known, they have the form of a differential geometry 
- the spacetime construct.  But in {\it Process Physics} we see a deeper understanding of reality in which  time  is not
geometry, it is
 process.


\begin{thebibliography}{99}

\bibitem{M} A.A. Michelson, {\it The Relative Motion of the Earth and the Luminiferous Aether}, 
{\it Amer. J. Sci}  {\bf S. 3} {\bf 22}, 120-129(1881).

\bibitem{MM} A.A. Michelson and E.W. Morley, {\it On the Relative Motion of the Earth and the
Luminiferous Aether},  {\it Philos. Mag S. 5}, {\bf 24}, No. 151,  449-463(1887).
 
\bibitem{RC02} R.T.  Cahill,   {\it  Process  Physics: From Quantum Foam to General Relativity}, 
gr-qc/0203015.

\bibitem{RC01} R.T.  Cahill,   {\it  Process  Physics: Inertia, Gravity and the Quantum}, to be
published in {\it Gen. Rel. and  Grav.}, gr-qc/0110117.

\bibitem{Aspect}  A. Aspect, J. Dalibard and G. Roger, {\it  Experimental Test of Bell's Inequalities Using Time- Varying
Analysers}, {\it Phys. Rev. Lett.} {\bf 49}, 1804-1807(1982).

\bibitem{Hardy} L. Hardy, {\it  Quantum Mechanics, Local Realistic Theories and
Lorentz-Invariant Realistic Theories}, {\it Phys. Rev. Lett.}, {\bf 68}, 2981(1992).

\bibitem{Percival}  I.C. Percival, {\it  Quantum Measurement Breaks Lorentz Symmetry}, quant-ph/9906005.

\bibitem{Munera}  H.A. Mun\'{e}ra, {\it Michelson-Morley Experiments Revisited: Systematic Errors,
Consistency Among Different Experiments, and Compatibility with Absolute Space}, {\it Aperion}
{\bf 5}, No.1-2, 37-54(1998).

\bibitem{Smoot}  G.F. Smoot {\it et al}, {\it Preliminary Results from the COBE Differential
 Microwave Radiometers - Large Angular Scale Isotropy of the Cosmic Microwave 
Background},  {\it  Astro. J., Part 2 - Letters}   {\bf 371}, April 10, L1-L5(1991).

\bibitem{BH}  A. Brillet and J.L. Hall, {\it  Improved Laser Test of the Isotropy of Space}, 
{\it Phys. Rev. Lett.} {\bf 42}, No.9, 549-552(1979).

\bibitem{Miller1}  D.C. Miller, {\it Ether Drift Experiments at Mount Wilson}, {\it Nat. Acad.
Sci.}  {\bf 11}, 306-314(1925).

\bibitem{Miller2}  D.C. Miller, {\it The Ether-Drift Experiments and the Determination of the
Absolute Motion of the Earth}, {\it Rev. Mod. Phys.} {\bf 5}, 203-242(1933).

\bibitem{MorMill}  E.W. Morley and D.C. Miller, {\it  An Experiment to Detect the
Fitzgerald-Lorentz Effect},  {\it Philos. Mag. S. 6}, {\bf 9},680(1905).


\bibitem{Illingworth}  K.K. Illingworth, {\it A Repetition of the M-M Experiment using Kennedy's
Refinement}, {\it Phys. Rev.} {\bf 30}, 692-696(1927).


\bibitem{CKK00}  R.T.  Cahill,  C.M.  Klinger,  and K. Kitto,    {\it Process Physics:
 Modelling Reality as Self-Organising Information}, {\it  The Physicist}   {\bf 37}(6), 191-195(2000),
gr-qc/0009023.

\bibitem{CK99} R.T. Cahill and  C.M. Klinger, {\it Self-Referential Noise
as a Fundamental Aspect of Reality},
 Proc.~2nd Int.~Conf.~on Unsolved Problems of Noise and Fluctuations
(UPoN'99), eds.~D.~Abbott and L.~Kish, Adelaide, Australia, 11-15th July
1999, {\bf Vol.~511,} p.~43 (American Institute of Physics, New York,  2000), gr-qc/9905082.

\bibitem{CK98} R.T. Cahill  and  C.M. Klinger, {\it Self-Referential Noise
and the Synthesis of Three-Dimensional Space}, {\it Gen. Rel. and  Grav.} {\bf 32}(3), 
529-540(2000), gr-qc/9812083.

\bibitem{MC} M.  Chown, {\it Random Reality}, {\it New Scientist}, Feb 26, {\bf 165}, No
2227, 24-28(2000). 


\end{thebibliography}
\end{document}